\documentclass[twocolumn]{aastex631}

\usepackage{svg}
\usepackage{amsmath}

\shorttitle{Evolution of LISA Observables for BBHs Lensed by an SMBH}
\shortauthors{Postiglione et al.}

\begin{document}

\title{Evolution of LISA Observables for Binary Black Holes Lensed by an SMBH}

\correspondingauthor{Jake Postiglione}
\email{jpostiglione@gradcenter.cuny.edu}

\author[0000-0003-0738-8186]{Jake Postiglione}
\affiliation{Department of Astrophysics, American Museum of Natural History, New York, NY 10024, USA}
\affiliation{Graduate Center, City University of New York, 365 5th Avenue, New York, NY 10016, USA}

\author[0000-0002-5956-851X]{K. E. Saavik Ford}
\affiliation{Department of Astrophysics, American Museum of Natural History, New York, NY 10024, USA}
\affiliation{Center for Computational Astrophysics, Flatiron Institute, New York, NY 10010, USA}
\affiliation{Graduate Center, City University of New York, 365 5th Avenue, New York, NY 10016, USA}
\affiliation{Department of Science, BMCC, City University of New York, New York, NY 10007, USA}

\author[0009-0009-6932-6379]{Henry Best}
\affiliation{Department of Astrophysics, American Museum of Natural History, New York, NY 10024, USA}
\affiliation{Graduate Center, City University of New York, 365 5th Avenue, New York, NY 10016, USA}
\affiliation{Department of Physics and Astronomy, Lehman College of the CUNY, Bronx, NY 10468, USA}

\author[0000-0002-9726-0508]{Barry McKernan}
\affiliation{Department of Astrophysics, American Museum of Natural History, New York, NY 10024, USA}
\affiliation{Center for Computational Astrophysics, Flatiron Institute, New York, NY 10010, USA}
\affiliation{Graduate Center, City University of New York, 365 5th Avenue, New York, NY 10016, USA}
\affiliation{Department of Science, BMCC, City University of New York, New York, NY 10007, USA}

\author[0009-0000-4476-5003]{Matthew O’Dowd}
\affiliation{Department of Astrophysics, American Museum of Natural History, New York, NY 10024, USA}
\affiliation{Graduate Center, City University of New York, 365 5th Avenue, New York, NY 10016, USA}
\affiliation{Department of Physics and Astronomy, Lehman College of the CUNY, Bronx, NY 10468, USA}

\begin{abstract}
Binary black holes (BBHs) are expected to form and merge in active galactic nuclei (AGN), deep in the potential well of a supermassive black hole (SMBH), from populations that exist in a nuclear star cluster (NSC). Here we investigate the gravitational wave (GW) signature of a BBH lensed by a nearby SMBH. For a fiducial GW150914-like BBH orbiting close to a $10^{8}M_{\odot}$ SMBH located at $z=0.1$, the lensed GW signal varies in a predictable manner in and out of the LISA detectability band and across frequencies. The occurrence of such signatures has the potential to confound LISA global fit models if they are not modelled. Detection of these sources provide an independent measure of AGN inclination angles, along with detecting warping of the inner disk, and measuring the SMBH spin.
\end{abstract}


\section{Introduction} \label{sec:intro}
The Laser Interferometer Space Antenna (LISA) is a European Space Agency (ESA) flagship mission to observe gravitational waves (GW) in the $f_{\rm GW} \sim 0.1-100$mHz frequency range \citep{LISA17,LISA23}, with an expected launch date in the mid-2030s. The mission will consist of three satellites forming a single triangular antenna. As a result, LISA will function as something of an `omni-scope' with all simultaneously detected GW convolved together, thus requiring sophisticated data analysis techniques to extract and distinguish signals from various astrophysical sources to create a catalog which can be used by electromagnetic (EM) observers \citep[e.g.][]{Littenberg13, Korol17,Sberna21}. The challenging task of de-convolving these overlapping signals can be eased by making some assumptions about the nature of the observed astrophysical sources and their expected associated waveforms \citep{LISA23}.

Many GW sources are expected to emit in the LISA frequency range, including massive black hole binaries (MBHB--spanning total mass $\sim 10^{4-7}M_{\odot}$ \citep{Sesana11,Mudit24}, white dwarf binaries (WDB) \citep{Breivik18,Littenberg20}, extreme mass ratio inspirals (EMRIs) \& intermediate mass ratio inspirals (IMRIs) \citep{Kocsis11,Amaro18,Derdzinski21,Peng23}, and of most importance to this work, stellar-mass binary black holes (BBHs) in the early stages of their inspiral \citep{Sesana16}. These BBHs are the same sources that are now routinely detected by higher frequency ground-based GW detectors (currently by the LIGO-Virgo-KAGRA [LVK] Consortium, e.g. \citep{GWTC3}). At the time of coalescence, BBHs are emitting at a strain sensitivity and GW frequency only detectable by ground-based GW detectors. However, a number of higher mass, nearby, BBH systems should appear in the LISA band years to decades prior to their detection by ground-based GW observers \citep[e.g.][]{Sesana16,Wagg22,Sberna22,McFACTS1}. Such potential multi-band sources open up a wide range of interesting astrophysical questions, which could be answered using coordinated observing campaigns. 

The following conventions are used to describe the parameters for a BBH system detectable by LISA: (1) $a_{\rm b,0}$, the semi-major axis of the binary at formation (if formed dynamically at smaller semi-major axis than appropriate to the LISA frequency range), (2) $\dot{a}_{\rm b}$, the decay rate of the semi-major axis (if purely GW driven or sped up through gas or dynamical interactions, e.g. $\ddot{a}_{\rm b}$), (3) ($a_{\rm b},e_{\rm b},i_{\rm b}$), the orbital parameters of the system appropriate to the LISA frequency band, including eccentricity around center of mass and inclination of orbit w.r.t. the observer's sight-line, (4) $dN_{\rm b}/dt$, the rate of occurrence of BBHs (which is related to both their detection rate by ground-based detectors but also the factors listed above).

Many authors have considered the LISA observability of BBHs which may form from dynamical channels, notably in globular clusters \citep[e.g.][]{Kremer18,Samsing18} or in Active Galactic Nucleus (AGN) accretion disks \citep[e.g.][]{McK14,Toubiana21,Sberna22}. The AGN channel of course rests on the assumption of a pre-existing nuclear star cluster (NSC) \citep[e.g.][]{Neumayer2020}. Isolated binary channels are also sources of LISA detectable BBHs \citep[e.g.][]{Wagg22,vanZeist23}, forming from stable mass transfer \cite[e.g.][]{Gallegos-Garcia_2021}, chemically homogeneous stellar binaries \cite[e.g.][]{10.1093/mnras/stw379}, and the stellar common envelope \cite[e.g.][]{Dominik2012ApJ}.  The original GW detection of a BBH, GW150914 \citep{150914} is often used as an example case, both because we know such systems exist, and if the system was evolving in vacuum, it \textit{would} have been detectable by LISA (had LISA been observing) $\sim$~a decade before its merger could be seen from the ground \citep{Sesana16}. If a GW150914-like BBH is orbiting close to a supermassive black hole (SMBH) in an AGN, we expect strong gravitational lensing effects. These effects have been infrequently considered in previous literature \citep[e.g.][]{Tamanini20,Chen:2020,Kuntz23,Zwick23}. Furthermore, it would not be unusual for such a binary to form in an AGN \citep{McFACTS1}.In the LVK band, the rate of occurrence for BBH mergers near the last migration trap is analytically estimated to be $\sim 0.4~\text{Gpc}^{-3}\text{yr}^{-1}$ for a $3 \times 10^8 M_\odot$ SMBH, where the AGN lifespan is $10^7~\text{yrs}$ \citep{Peng21}. We will analyze the uncertainties in this rate in our discussion section. While less likely, it is also possible for such a system to form in a gas-poor NSC, where it could experience similar lensing effects.

Gravitational lensing occurs when an object of sufficient mass comes between a GW source and an observer. The lens magnifies the emitted signal, increasing the apparent luminosity of the source. This effect is observed in EM, but should also occur in GW. For LVK mergers, this lensing would cause the binary to appear as high-mass and at a small redshift, but current observations have not confirmed such an event \citep{Chen:2020}.

Here we consider the simple case of a BBH orbiting the SMBH on a small, circular, co-planar orbit at 2 different radii within an AGN disk. We further assume the BBH itself has an eccentricity of zero ($e_{\rm b}=0$) with respect to its own center of mass, consistent with the observed eccentricity of GW150914. We ignore gas and tidal effects on the BBH's evolution, simply assuming vacuum evolution of $a_{\rm b}$ to examine the impact of lensing on the LISA-observed $f_{\rm GW}$ and strain ($h$) from such a source. We invoke AGN channel models for the formation of this hypothetical binary, finding that there is a potential to form a binary very close to the SMBH if multiple single black holes are efficiently delivered to the inner disk \citep[e.g.][]{Secunda21}. \cite{Bellovary16,Peng21} have also shown that migration traps may exist at small disk radii, depending on the model \citep[e.g.][]{Sirko03}, although also see \cite{Grishin24}. For disks close to edge-on to the observer, BBHs at small radii can experience extremely strong gravitational lensing effects from the SMBH, along with strong red-shifting due to the deep gravitational potential. The observed GW waveforms should depart substantially from vacuum expectation and will show strongly perturbed observed strain and frequencies, which will vary with the phase of the BBH's orbit around the SMBH \citep{Kuntz23}. The BBH's orbital timescale around the SMBH will be much smaller than the LISA mission lifetime, so these signatures varying periodically in ($f_{\rm GW}, h$) must be accounted for if such a source is to be identified in the LISA data stream.

\begin{figure}[t!]
    \centering
    \includegraphics[width=\columnwidth]{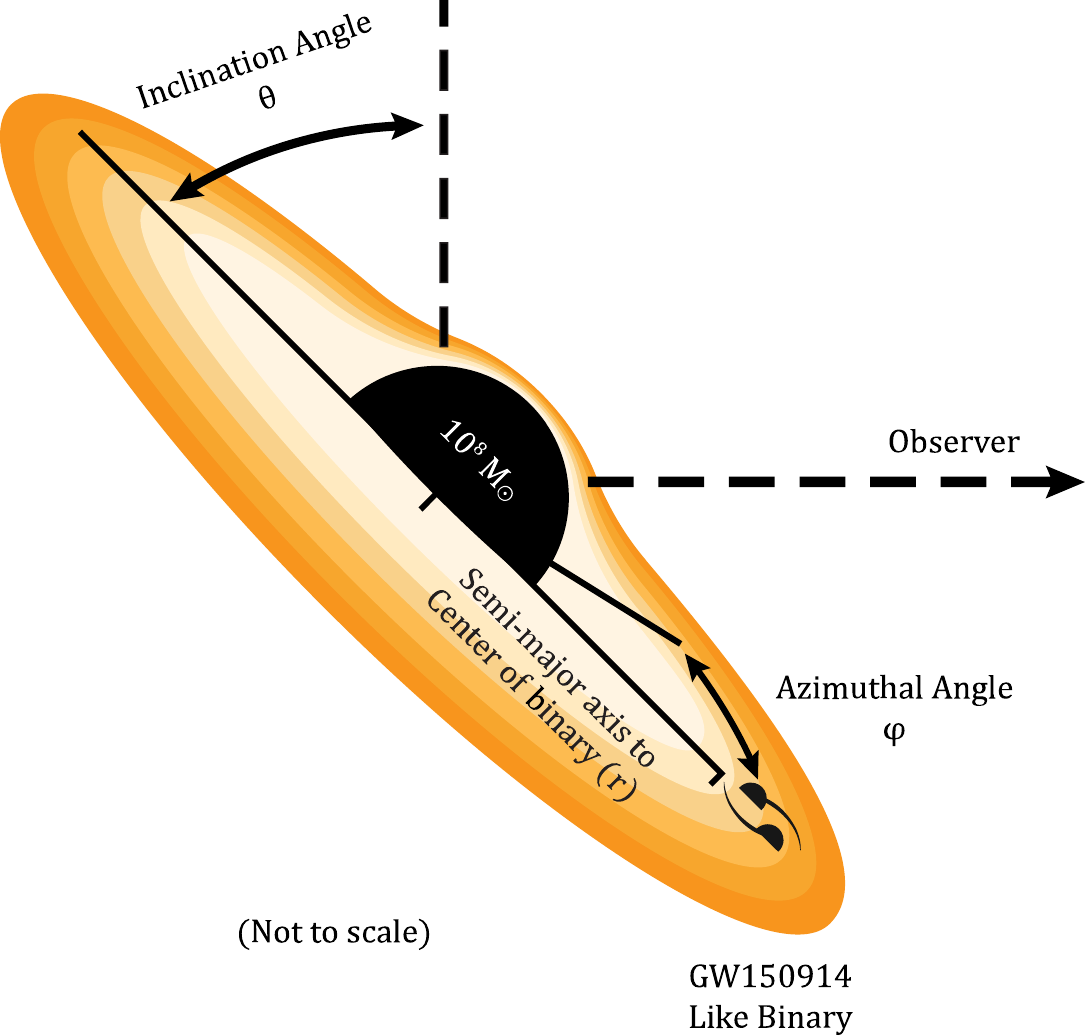}
    \caption{Illustration showing the physical setup for the environment in which we consider a GW150914 like binary to be in orbit around a $10^8~M_\odot$ SMBH. The center of mass of the binary is tracked in polar coordinates. The semi-major axis of the binary's orbit is measured in gravitational radii $R_g$. The azimuthal angle $\varphi$ of the binary in the disk is measured from the axis extending to the observer. The inclination $\theta$ of the disk is measured from being face-on to the observer, with the top of the disk moving away from the observer as the inclination increases.}
    \label{fig:phys-setup}
\end{figure}

\section{Methods} \label{sec:methods}
We consider a binary system of the same description as GW150914 with masses of $36~M_\odot$ and $29~ M_\odot$ placed at a distance of $410$~Mpc from a theoretical observer \citep{PhysRevLett.116.061102}. Along with the BBH being nearly circular, previous studies focusing on the possible multi-band detectability of GW150914 make it an obvious fiducial system to base our tests on. The system would have had an orbital frequency of $f_{orb} = 0.8 \times 10^{-3}$~Hz roughly 5 years before LIGO detected its coalescence, and detectable by LISA had it been observing \citep{Sesana16}. We consider this binary in a circular, co-planar orbit, without perturbation, around an SMBH. The mass of the SMBH was chosen to be $10^8~M_\odot$ to artificially limit the effects of tidal disruption on the binary, if a full numerical simulation was run (see below).

Figure \ref{fig:phys-setup} shows the physical setup of our model and how the system is oriented with respect the observer. We assume the binary to be in a circularized orbit, so the position of the binary's center of mass is tracked in polar coordinates. The orbital radius $r$ around the SMBH is measured in units of gravitational radii $R_g = GM_{\rm SMBH}/c^2$. The azimuthal angle is given as $\varphi$, where an observer exists along the axis extending from $\varphi = 0^\circ$. The inclination angle of the disk $\theta$ is defined such that a disk with an inclination of $\theta = 0^\circ$ would be face-on to the observer. We pick inclination angles of $\theta=88^\circ$ and $\theta=65^\circ$, so we can explore one highly inclined case and one less inclined case, respectively. We further discuss the range of inclination angles, for which our results are valid, in section \ref{sec:discussion:rates} below. We can also specify the semi-major axis of the (presumed quasi-circular) BBH as $a_\textrm{b}$ (not pictured), as it evolves over time. Throughout the work, we assume an initial semi-major axis of $a_\textrm{b} = 1 \times 10^{-3}$~AU, which corresponds to an GW emission frequency of $f_{\rm{obs}} = 1.6 \times 10^{-3}$~Hz. 

Due to the extremely small values of $r$ we consider in this work, it is worth checking that our proposed binary is stable to tidal forces from the SMBH. If we compute the value of the Hill Radius, $R_{\rm Hill} = r (q/3)^{1/3}$, where $q$ is the small mass ratio between the binary and the SMBH, for $r=20~R_g$, we find $R_{\rm Hill}>a$; were we to substitute a smaller SMBH mass of $10^6 M_{\odot}$ (where our proposed binary might also provide a very interesting LISA-detectable EMRI signal) we would unfortunately have $a>R_{\rm Hill}$. Thus, despite the possibly interesting nature of such a source, modeling it would require a more sophisticated approach and is beyond the scope of this work.

\subsection{Ray tracing}
The gravitational potential surrounding a central SMBH is typically described using either the Schwarzschild or the Kerr metric. Assuming that gravitational waves propagate along null geodesics (e.g., like photons), we make use of preexisting ray-tracing solutions. In particular, we use \texttt{Sim5}\footnote{https://github.com/mbursa/sim5/} \citep{2017bhns.work....7B} to trace geodesics from the observer to the mid-plane of an thin inclined accretion disk. \texttt{Sim5} assumes a Kerr metric; however, for simplicity, we set the spin of the SMBH to 0. This results in a Schwarzschild metric, meaning our results only apply to cases where the SMBH is not spinning. Since a spinning SMBH is well motivated, it would be worth repeating our investigation for a variety of SMBH spins in future work.

Certain regimes of parameter space require proper treatment of gravitational wave propagation using wave optics~\citep{Dorazio20, Dalang22}. 
There are three major sources of interference which we should consider: scattering, magnification, and the modulation of the wave's polarization.
If we consider the scattering of gravitational waves by a third body (e.g. the SMBH), we must determine how strong the scattered waves appear to the observer and if they may considerably interfere with the original source~\citep{Dolan08, Pijnenburg24}.
As described in~\citet{Dorazio20}, wave optics become important in this type of system when the wavelength of the emitted GWs is longer than the Schwarzschild radius of the third black hole, which is the case when $f_{\rm{GW}} \geq 0.08 \rm{Hz} \left(10^{5} M_{\odot} / M_{\rm{BH}} \right)$.
This effect does not require traditional lensing, and will occur at any point of the orbit the BBH.
For $f_{\rm{GW}} = 1.6 \times 10^{-3} \rm{Hz}$, this corresponds to SMBH with $M_{\rm{BH}} \leq 10^{6} M_{\odot}$, and our experiment is safely outside this range due to the large mass of the SMBH.

Wave optics may also start becoming important in the context of gravitational lensing of the BBH system.
To estimate the first order correction to geometric optics, we follow section 3 and 4 of~\citet{Dalang22} to determine the magnitude of distortions of the gravitational wave\footnote{Note that this is a first-order approximation for a point-like lens.}:
\begin{equation}
\label{equation_gravitational_wave_distortion_by_big_bad_smbh}
\begin{split}
    F_{\mu\nu\alpha\beta} &= \left[m_{\mu}m_{\nu} - i \frac{4\Omega R_{\rm{s}}}{b^{2}}e^{2 i \delta} n_{\mu} n_{\nu}\right] l_{\alpha} l _{\beta} \\
    &\quad + \left[l_{\mu}l_{\nu} - i \frac{4\Omega R_{\rm{s}}}{b^{2}}e^{-2 i \delta} n_{\mu} n_{\nu}\right] m_{\alpha} m _{\beta} ,
\end{split}  
\end{equation}
where $i\equiv\sqrt{-1}$ the imaginary unit, $\Omega \equiv 2\pi / \lambda_{\rm{emit}}$ the energy of the wave, $\lambda_{\rm{emit}}$ is the emitted wavelength from the GW source, $R_{\rm{s}}$ is the Schwarzschild radius of the SMBH, $b$ is the impact parameter often measured in units of $R_{\rm{g}}$, $\delta$ is the angle between the plus polarization and the impact position in the lens plane, and $n_{\mu}, m_{\mu}, \text{ and } l_{\mu}$ are basis vectors such that in the unperturbed metric: $\bar{n}^{\mu}\propto (1, 0, 0, -1), \bar{m}^{\mu}\propto (0, 1, i, 0), \text{and } \bar{l}^{\mu} = \bar{m}^{*\mu}$. 
To first order, the perturbed components become: $n^{\mu}\propto (1, 0, 0, -1), m^{\mu}\propto ( -R_{\rm{s}} b^{-1}e^{i\delta}, 1, i, R_{\rm{s}} b^{-1}e^{i\delta}), l^{\mu}\propto (-R_{\rm{s}} b^{-1}e^{-i\delta}, 1, -i, R_{\rm{s}} b^{-1}e^{-i\delta})$.
In equation (\ref{equation_gravitational_wave_distortion_by_big_bad_smbh}), the first set of terms in the brackets represent the undistorted tensor where geometric optics hold.
Importantly, the second term in each set of brackets represents deviations from the geometric optics regime, which scales $\propto R_{\rm{s}} \lambda_{emit}^{-1} b^{-2}$.
Within our experiment, the configuration most likely to lay outside the geometric optics regime is the inclined accretion disk at 88$^{\circ}$ with the BBH orbiting the SMBH at 20 $R_{\rm{g}}$. 
In this orbit, when the BBH is farthest from the observer, the impact parameter becomes $b = 20 R_{\rm{g}} \cos{88^{\circ}} \sim 0.7 R_{\rm{g}}$ (the impact parameter does not depend on the curvature of the metric).
The redshifting of the GW by the SMBH's gravitational well and orbital motion should be considered, which is approximately 0.5 when the BBH is 20 $R_{\rm{g}}$ from the SMBH and appears to be receding from the observer (e.g. see Figure~\ref{fig:radii-interp}). We consider this by approximating the wavelengths observed by LISA as $\lambda_{\rm{emit}} = \lambda_{\rm{obs}} (1 + z)^{-1} = c f_{\rm{obs}}^{-1}(1 + z)^{-1} \approx 10^{-11} \rm{m}$.
From equation (\ref{equation_gravitational_wave_distortion_by_big_bad_smbh}), we determine that the magnitude of distortion is approximately $2 \pi R_{\rm{s}} \lambda_{\rm{emit}}^{-1} b^{-2}$, and with relevant values becomes $ \approx (10^{11} \rm{m}) (10^{-11}\rm{m}^{-1}) (2) \approx 2$. 
We note that this reduces to distortions that are $\approx b^{-2}$, meaning this is in the wave optics regime and our geometric solution does not truly represent the expected modulation of signal.
However, this is an evolving system so we should determine if there is a significant amount of time that we are within the wave optics regime.
This may be approximated as the arc length of the orbit around the SMBH where the GW's tensor is strongly affected by lensing.
This may be estimated as when the correction terms in equation (\ref{equation_gravitational_wave_distortion_by_big_bad_smbh}) approach a considerable fraction of unity.
We define this threshold to be 0.1, defined as the BBH's position with an impact parameter $b < \sqrt{10}R_{\rm{g}}$, representing a 10\% deviation from geometric optics.
For an orbit with radius $R_{\rm{orbit}}$ in units of gravitational radii, this is only a region within $\approx 9^{\circ} = \arccos\left(\sqrt{(1 - 10R^{2}_{\rm{g}}/R_{\rm{orbit}}^{2}) / \sin^{2}(\theta)}\right)$ behind the SMBH.
This limit is never reached in the configurations with $\theta = 65^{\circ}$ or $R_{\rm{orbit}} = 100 R_{\rm{g}}$, and accounts for only $\sim$5\% of the orbit in the configuration $\theta = 88^{\circ}, R_{\rm{orbit}} = 20 R_{\rm{g}}$.
Therefore, for our purposes of order of magnitude estimations across the 4 years of observations, we may use geometric optics to approximate the redshifting and amplification due to gravitational lensing.

Wave optics may also be required to determine the change of the GW's polarization.
The typical plus and cross polarizations have the ability to interfere with each other if they remain in phase, become circularly polarized if one path around the strong lens accrues a half wavelength phase shift, or a combination of the previous~\citep{Pijnenburg24}.
This is most important in unresolved multiply imaged systems such as traditional strong lensed systems~\citep[see][for a review on lensing]{Courbin02}.
However, the creation of multiple discrete images requires crossing a \emph{caustic} in the source plane, a region where the magnification formally diverges and leads to the creation or destruction of images.
A caustic cannot be produced by a single SMBH, and requires some external shear or imperfect lensing potential.
Therefore, we do not consider these effects, as we only are looking at the order of magnitude amplification of the BBH's signal as it orbits around an SMBH.
We note that this may be an important consideration in cosmological scale lenses where a galaxy's imperfect Fermat potential leads to the splitting of images.

The projection we use simulates how the accretion disk might appear to an observer after considering light bending effects, if we were able to resolve the $\sim$nanoarcsecond angular structure.

Our geodesics define test locations on the accretion disk where the generalized Doppler factor $g \equiv f_{\rm{obs}} / f_{\rm{emit}} = 1/(z + 1)$ and spatial positioning, important for the calculation of magnification, are recorded. The generalized Doppler factor from this simulation is a sum of both the velocity and gravitationally dependent redshifting a photon emitted from the disk will exhibit.

By considering the positioning of geodesics in the accretion disk and the energy shift of its radiation, we can predict the observed frequency and magnification as functions of time for an evolving binary emitting gravitational waves that are orbiting in the mid-plane.

We set up our ray-tracing simulation by placing the SMBH at the center of the screen and preparing a grid of geodesics which connect the observer to the plane of the accretion disk. This grid extends out 200 $R_{\rm{g}} \equiv GM_{\rm{SMBH}}/c^{2}$ in each direction, leading to a 400 by 400 $R_{\rm{g}}$ square.

In the case of inclined disks, positive y-values represent looking ``over'' the SMBH at the far side of the disk. This region of the accretion disk typically appears magnified due to gravitational lensing. It is important to note that, due to light bending effects, the incident shape of the grid onto the inclined disk strays from the its original square-like configuration.

The radial position $r_{\rm{emit}}$, the azimuthal angle $\phi_{\rm{emit}}$, and the generalized Doppler factor $g(r_{\rm{emit}}, \phi_{\rm{emit}})$ for where a geodesic crosses the mid-plane, correspond to the geodesic's originating point in the grid.

\subsection{Redshift and Magnification Interpolation}

\begin{figure}[b!]
    \centering
    \includegraphics[width=0.5\columnwidth]{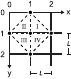}
    \caption{Showing the extents of the triangles used to calculate the average magnification for a point at (1, 1). The grid corresponds with the starting position of each geodesics indexed by $x$ and $y$, separated by length $L$.}
    \label{fig:tri-grid}
\end{figure}

\begin{figure*}[t!] 
    \includegraphics[width=\textwidth]{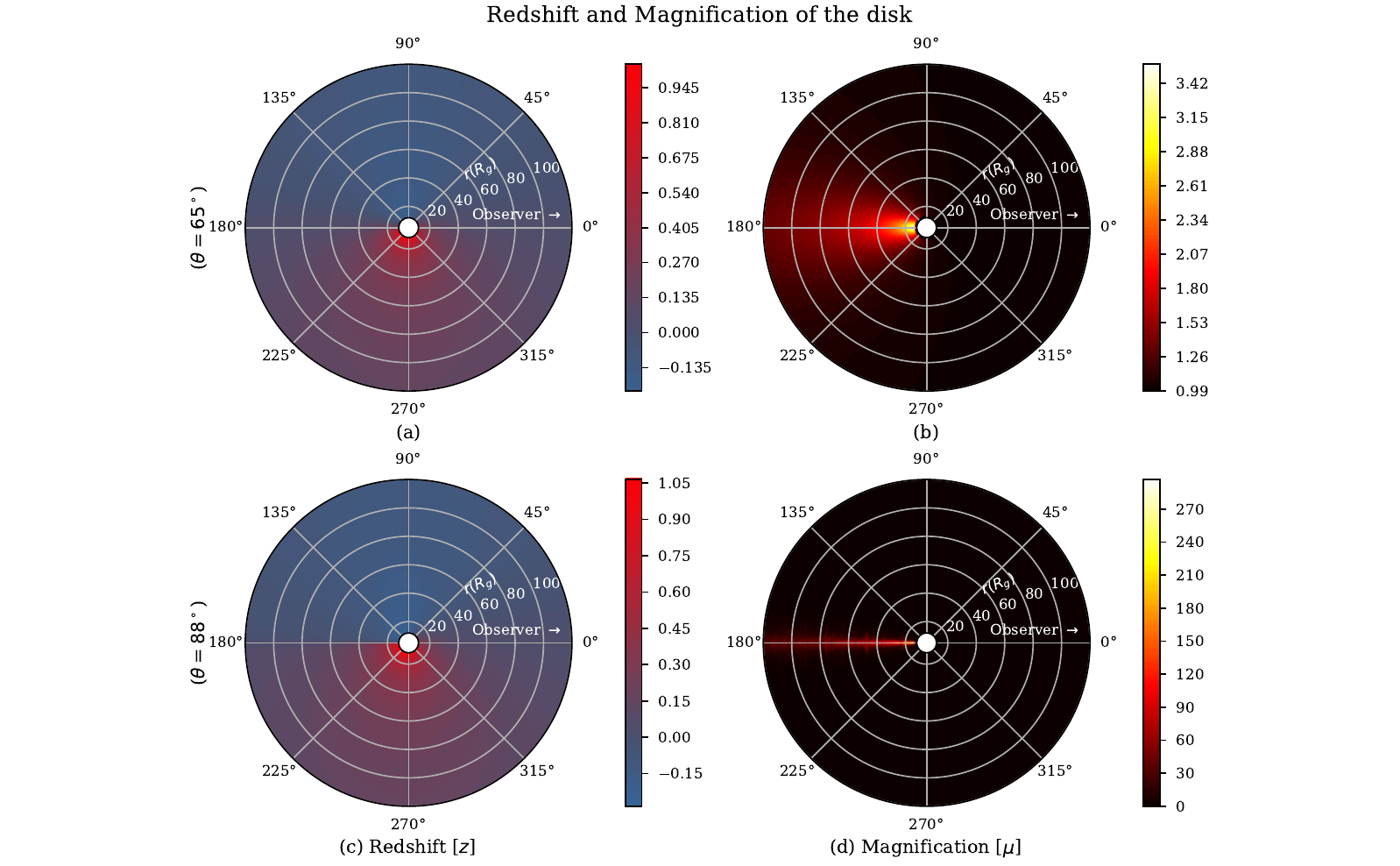}
    \caption{Contour plots showing the redshift and magnification sampled across the disk ranging from $12R_g$ to $120R_g$ for two separate inclination angles of $65^{\circ}$ and $88^{\circ}$. The observer is looking at the disk from an azimuthal angle of $\theta = 0^{\circ}$ and the inclination angle is defined such that the $\varphi = 0^{\circ}$ case is face-on. At higher inclination angles, the top of the disk tilts away relative to the observer's perspective. Note the difference in scales between the $65^{\circ}$ and $88^{\circ}$ cases for both redshift and magnification.}
    \label{fig:disk-interp}
\end{figure*}

Due to the grid-like structure of the data given by our simulation, it is necessary to interpolate along the orbital path for the redshift and magnification. We utilize the \texttt{griddata} function from \textit{SciPy} to interpolate over the data using test points along an orbit with a radius defined by $n$ gravitational radii. As discussed above, redshift is calculated by the simulation, but magnification requires we compare the position of the starting and ending positions of the geodesics. We define this magnification factor as $\mu = A_{\rm{obs}} / (A_{\rm{disk}} \cdot cos \theta_i)$ where $A_{\rm{obs}}$ is the initial area of a triangle formed by three neighboring origin points of geodesics in the grid, and $A_{\rm{disk}}$ is the area of a triangle formed by the same geodesics but instead using the disk-crossing position. Figure \ref{fig:tri-grid} shows how the initial triangles are defined using the grid position of geodesics as they are aligned with an integer coordinate grid. The average of four triangles, each respective to a quadrant around an integer coordinate, is used to find the final magnification factor $\mu$ for one a geodesic. There exist edge cases at the beginning and end of a column or row, where less than four triangles are used to find the average magnification; however, the edges of the grid fall outside of the area we use in our experiment.

We display the resulting maps, zoomed in to range from 12 to 120 $R_g$, in Figure \ref{fig:disk-interp}. Plots (a) and (c) show the interpolated redshift for an inclined disk at $\theta=65$ and $88$~degrees respectively. Plots (b) and (d) show the interpolated magnification for these respective inclinations. Note the difference in scale between the $65^{\circ}$ and $88^{\circ}$ cases for both redshift and magnification.

\begin{figure*}[t!] 
    \includegraphics[width=\textwidth]{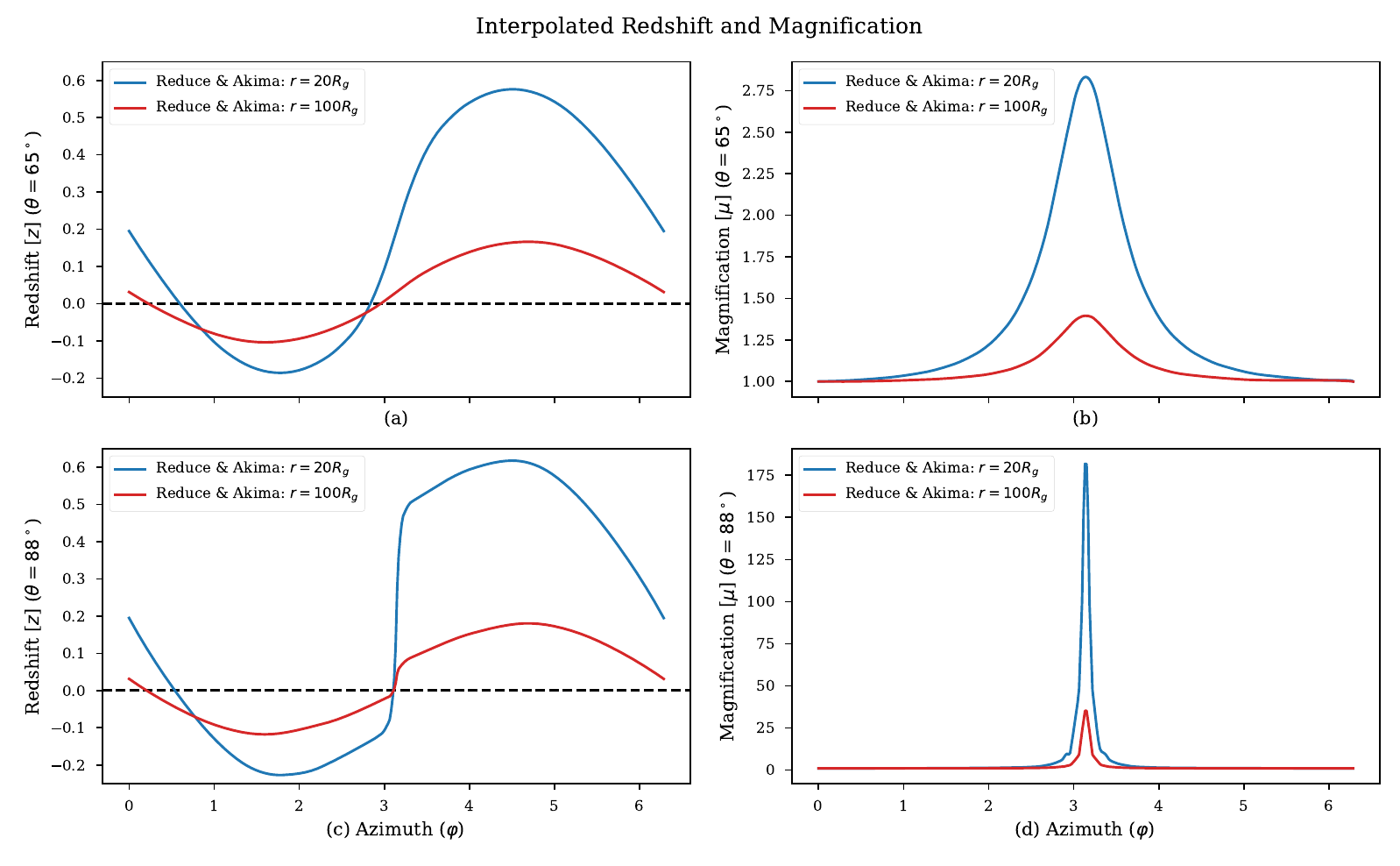}
    \caption{Plots showing the interpolated change in redshift and magnification vs. azimuthal angle in the disk at gravitational radii of $20R_g$ and $100R_g$ for separate inclination angles of $65^{\circ}$ and $88^{\circ}$. Note the difference in maxima for the magnification between the two cases.}
    \label{fig:radii-interp}
\end{figure*}

We consider the binary at both $20~R_g$ and $100~R_g$, near where migration traps are expected to exist, given favorable disk morphology \citep[][]{Bellovary16, Peng21}. Figure \ref{fig:radii-interp} shows the interpolated values for the redshift and magnification as as function of the azimuthal angle ($\varphi$), note the difference in maxima of the interpolated magnification between the $65^{\circ}$ and $88^{\circ}$ disks. The interpolation is achieved by first reducing the number of co-linear points in the data and then applying an Akima spline, allowing for the interpolation of values while preserving important features \citep{akima1970}. It is important to note that the peak of the magnification occurs when the redshift is changing \textit{most} rapidly in time, along with the differing in shape of the data and scale of the vertical axis between the different inclination angles.

\subsection{Binary Evolution}

The orbital period $P$ for a binary orbiting around an SMBH at radius $r$ in units of $R_g$ is given by $P(r)=2\pi G M_{\rm{SMBH}} r^{3/2} c^{-3}$. The azimuth at which we sample our interpolated redshift and magnification can be found as $\varphi = (2 \pi T P^{-1})\bmod{2\pi}$ where $T$ is the total elapsed time.

We utilize \texttt{LEGWORK}, an open source project developed to predict the evolution of binary sources due to gravitational wave emission \citep{LEGWORK_joss, LEGWORK_apjs}, producing a baseline for the evolving observed frequency, strain, and characteristic strain. To find the magnified strain amplitude $h_{\mu}$, the strain amplitude $h_0$ is multiplied by the square of the magnification factor $\sqrt{\mu}$, given by the position of the binary along its orbit. The strain amplitude $h_0$ is obtained via \texttt{LEGWORK} for a theoretical observer at a distance of $410$ ~Mpc.

However, this does not account for the fact that the frequency seen by a distant observer is changing rapidly, which changes the \textit{characteristic strain} we would expect to see in the LISA signal, and the inferred SNR for such a source in the LISA data stream. Assuming the system to be circular, isolated, and unperturbed by gas drag or other external dynamical forces, we can approximate the characteristic strain as $h_c^2 = \frac{N}{2^{5/3}} h_\mu^2$ where $N \approx (f_{\rm obs}^{2} / |\dot{f}_{\rm obs}|)$ over the period of its evolution through the LISA frequency regime. \footnote{Since we use the strain amplitude given by LEGWORK, we must account for an extra $2^{5/3}$ resulting from eq. 17, 26, and 27 in \citet{LEGWORK_joss, LEGWORK_apjs}.}

\section{Results} \label{sec:results}

\begin{figure*}[t!] 
    \includegraphics[width=\textwidth]{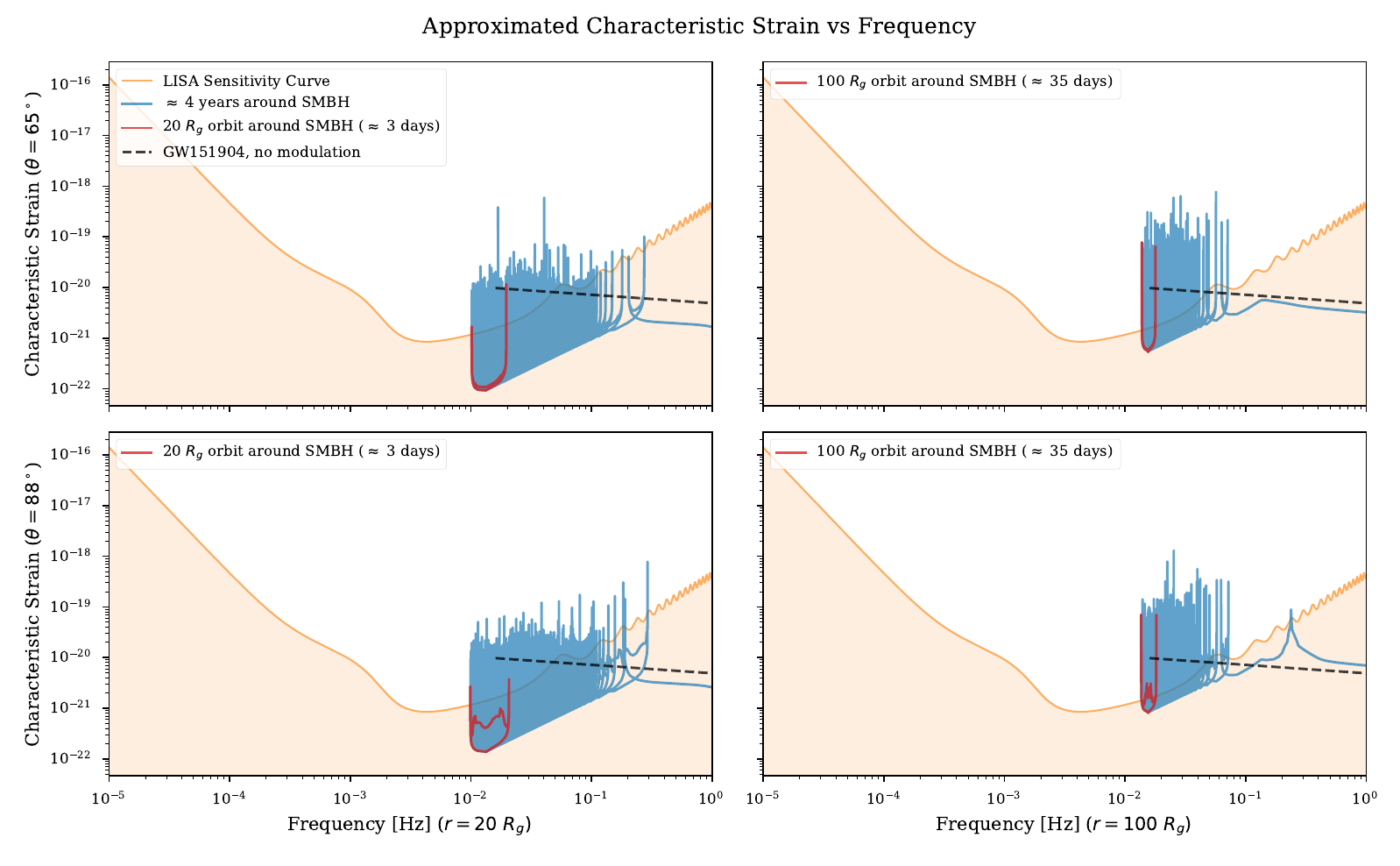}
    \caption{Characteristic Strain vs. Frequency for a binary at $20R_g$ and $100R_g$ in disk with inclinations of $65^\circ$ and $88^\circ$. The two peaks correspond to the where the rate of change of the redshift is small. The minima of the evolution corresponds to when the binary is positioned directly in front of or behind the SMBH, where magnification is at a minimum or maximum, respectively. The $20R_g$ orbit would have a period of $\approx \text {3 days}$, with the $100R_g$ orbit having a period of $\approx \text {35 days}$. This would correspond to a burst-like signal occurring every 1.5 and 17.5 days respectively.}
    \label{fig:approx-h-grid}
\end{figure*}

Using the interpolated redshift and magnification, shown in Figure \ref{fig:radii-interp}, to approximate the characteristic strain throughout the evolution of the binary allows us to plot the characteristic strain vs. frequency over the LISA sensitivity curve, as seen in Figure \ref{fig:approx-h-grid}. The evolution of the isolated binary in a vacuum and assuming pure GR evolution, can be seen in magenta. The first orbit of the binary around the SMBH can be seen in orange, with the entire lifespan of the binary in the LISA frequency range seen in blue.

Since the characteristic strain partially depends on the observed change in gravitational wave frequency, the peaks occurring twice an orbit in Figure \ref{fig:approx-h-grid} do not correlate to the peak in magnification seen in plots (b) and (d) of Figure \ref{fig:radii-interp}. When the binary passes behind the SMBH, relative to the sight line of the observer (around an azimuth of $\varphi \approx \pi$ in Figure \ref{fig:radii-interp}), the observed strain amplitude is highly magnified. However, the frequency is quickly changing ($\dot{f}_{\rm obs}$ is large) due to the rate of change of the redshift. In effect, for both disk inclinations and orbits, the characteristic strain is significantly decreased. In the 20~$R_g$ case, this causes the signal to fall below below the LISA sensitivity curve for both inclination angles. 

The dual peaks observed in Figure \ref{fig:approx-h-grid}, therefore correspond to portions of the orbit where there is \textit{some} magnification and the change in redshift/frequency is slowest ($\dot{f}_{\rm obs}$ is near zero). This can be seen around an azimuth of $\varphi \approx \pi/2$ and $\varphi \approx3\pi/2$ in Figure~\ref{fig:radii-interp}.

We note several additional important trends visible in Figure~\ref{fig:approx-h-grid}: going from $r=20$ to 100~$R_g$ we see the characteristic strains are larger at the same frequencies since the change in $\dot{f}_{\rm{obs}}$ is slower as a result of a larger, slower orbit. Going from an inclination of $\theta=65$ to $88$~degrees, the trace of the plot during an orbit is different since the magnification factor is higher when the binary is behind the SMBH in the more inclined disk. This feature is present in both $r=20$ and 100~$R_g$ orbits. Finally, in all cases we note the sharp burst-like quality of the observed lensed signals in the higher frequency region of the LISA band, where background sources are expected to be less common. We will return to this latter point in the discussion section.

\section{Discussion} \label{sec:discussion}

\subsection{Rates and LISA Detectability} \label{sec:discussion:rates}
In preparation for when LISA starts observing, we want to build a more complete picture of our universe so that we can either confirm the accuracy existing models, or know that new theories need to be developed. To start in this endeavor, we can look at existing rate calculations for the LVK population.  

The occurrence of BBH mergers near the last migration trap is analytically estimated to be

\begin{equation}
    R= \frac{n_{\rm QSO}}{2} \frac{N_{\rm BH}}{\tau_{\rm AGN}} \sim 0.4~\text{Gpc}^{-3}\text{yr}^{-1}
\end{equation}
where $n_{\rm QSO}=10^{-7}~\text{Mpc}^{-3}$ is the number density of QSOs, $N_{\rm BH}=6 \times 10^4$ is the number of black holes in the NSC, and $\tau_{\rm AGN}=10^7~\text{yrs}$ is the assumed AGN lifetime \citep{Peng21}. However, this estimate implicitly assumes an aggressive prescription for the disk capture rate of BH from the NSC---nearly all of the BH are captured by the end of the AGN phase. On the other hand, it also assumes a conservative number density of QSO (only the most luminous) which does not account for the full spectrum of BBH forming QSOs and AGN (excluding lower mass \& luminosity) that could exist in a detection volume out to $z \simeq 1$. These choices work out such that assuming the opposite (a less aggressive capture rate, but high number density of AGN) could yield a similar answer; additionally the AGN lifetime is sufficiently uncertain that it contributes to uncertainty in our rate estimate and we could imagine a rate that varies by an order of magnitude (or more) in either direction.

One way to both test and extend this estimate for LISA would be to utilize a population synthesis code such as \textit{McFACTS} \citep{McKernan:2024kpr}. Depending on the properties of the AGN disk used to run \textit{McFACTS}, one would be able to count the occurrence of objects fitting our criteria and assume a rate at which we expect LISA to observe these types of objects.
\textit{McFACTS} currently does not permit binary formation in the inner disk (even well outside the last migration trap), however we can determine a rough estimate for the rate by noting that in a typical run, $\sim2\%$ of all BH end up in the inner disk and could form binaries at the last migration trap \citep[e.g.][]{McFACTS1}. Assuming \textit{all} BBH mergers originate from the AGN channel, we would expect at most $2\%$ of the LVK merger rate could come from these strongly lensed binaries or $\sim0.5~\text{Gpc}^{-3}\text{yr}^{-1}$. This is notably similar to the rate from \citet{Peng21}, but as with their estimate, there remain at least order of magnitude uncertainties.

One thing to consider about the detectability of these types of sources in LISA, is the range or threshold of inclination angles of the disk where we might reasonable expect the change in observed frequency due to Doppler shifting to become larger than width of the frequency bin after some observing times. The width of a frequency bin for LISA is described as the inverse of the total observing time in seconds. For one year, the width of a single bin is expected to be $\sim 3.169 \times10^{-8}~ \text{Hz}$. As such, for LISA to observe an apparent periodic change in frequency, the delta between the extrema of the frequency must be larger than the bin width. Keeping the same assumptions of the system we make earlier, we can approximate the critical inclination angle $\theta_{\text{crit}}$ at which we would expect the change in frequency to cross into multiple bins:
\begin{equation}
    \theta_{\text{crit}} = \arcsin\left(\frac{c}{v_{\text{orb}}} \frac{\alpha - 1}{\alpha + 1}\right)
\end{equation}
where $\alpha = \left(\frac{1}{2 f_{\text{emit}} T_{\text{obs}}} + 1\right)^2$, $v_{\text{orb}}$ is the orbital velocity, $f_{\text{emit}}$ is the emitted GW frequency of the black hole binary, and $T_{\text{obs}}$ is the total observational run time in seconds. For an observational run of 1 year, a binary with a circular orbit at $20~R_g$, emitting at a frequency detectable by LISA, would have an detectable frequency shift over multiple bins for disks with inclination angles over $\sim 4.43 \times 10^{-6} ~\text{rad}$. Similarly, for a binary at $100~R_g$, this would occur in disks with inclination angles over $\sim 9.90 \times 10^{-6} ~\text{rad}$. For the $100~R_g$ case, and assuming a uniform distribution of inclination angles ranging from 0 to $2 \pi$, only $\sim 6.3\times10^{-4}~\%$ of disks would have inclinations smaller than the critical angle $\theta_{\text{crit}}$.

\subsection{A burst, or not a burst, that is the question}
Depending on where these sources are in the disk, the observed characteristic strain would spend most of its time below the LISA sensitivity curve (notably and perhaps counterintuitively, this is worse for sources closest to the SMBH, which experience the largest accelerations and therefore suffer from the fastest changing observed frequencies). This may lead to their appearance as ``bursty'' sources, similar to those expected from \citet{Knee_2024}. However, unlike the bursts defined in \citet{Knee_2024}, these sources may be made more detectable by summing the repeated strain contributions to a particular frequency bin, though the intrinsic evolution of the binary will counteract this. We also note that the ``bursts`` we expect from this type of source occur at relatively high frequencies within the LISA band, meaning there should be fewer alternative astrophysical sources at the relevant frequencies and strains \citep{colpi2024lisadefinitionstudyreport}. In principle (as seen in Fig.~\ref{fig:approx-h-grid}, the sources \textit{are} detectable at their peak characteristic strains, so the problem becomes not one of simple detection, but one of distinguishability---i.e. can we recognize the source of the signal as a real astrophysical source, and determine the astrophysical origin of the signals we observe. The strategies proposed by \citet{Knee_2024} should help with this problem.

\subsection{Intrinsic Properties of the Disk}
Detecting a binary as we described, in an AGN disk, could provide an extremely useful probe of the underlying disk morphology. In our simplified case, it is apparent that one would be able to infer the inclination of the disk (at the orbit of the binary) based on the change in strain amplitude between the binary passing in front of and behind the SMBH. A larger change in strain would be produced by a more edge-on orientation (as is apparent by comparing the lower panels to the upper panels in Fig.~\ref{fig:approx-h-grid}. On the other hand, a precisely face-on orientation would produce no magnification or time-varying Doppler shift (though there would still be a constant frequency offset from the emitted to observed frequency, due to gravitational redshifting). If the binary is sufficiently close to the SMBH, we would expect the emitted gravitational waves incur effects from relativistic beaming. In such a case, we would expect to see a difference between the local maxima of the two peaks of the characteristic strain, therefore one could determine the phase alignment between the GW observables and other possible observables of the system. 

\begin{acknowledgments}
Support was provided by Schmidt Sciences, LLC. for JP, HB \& MO. KESF \& BM are supported by NSF AST- 2206096, NSF AST- 1831415 and Simons Foundation Grant 533845 as well as Simons Foundation sabbatical support. The Flatiron Institute is supported by the Simons Foundation. HB acknowledges the support from the Czech Science Foundation (GACR) Junior Star grant no. GM24-10599M. KESF wishes to acknowledge extremely helpful conversations with Xian Chen and Laura Sberna at \textit{New Ideas on the Origin of Black Hole Mergers} at NBIA, in August 2024. JP, KESF, and BM wish to acknowledge the useful conversations and abundance of help provided by Tom Wagg throughout this exploration. The authors also wish to acknowledge the helpful suggestions and advice provided by the referee in their evaluation of the manuscript of this article.
\end{acknowledgments}

\software{Astropy \citep{astropy:2013, astropy:2018, astropy:2022}, LEGWORK \citep{LEGWORK_joss, LEGWORK_apjs}, Matplotlib \citep{Hunter:2007}, Numpy \citep{2020NumPy-Array}, SciPy \citep{2020SciPy-NMeth}, SIM5 \citep{2017bhns.work....7B, 2018ascl.soft11011B}}

\bibliography{main}{}
\bibliographystyle{aasjournal}

\end{document}